\documentclass[twocolumn,prl,showpacs]{revtex4}

\usepackage{graphicx}
\usepackage{dcolumn}
\usepackage{bm}
\usepackage{amsmath}

\begin{document}

\preprint{Helical edge}
\title{Surface  density of states and topological 
edge states in non-centrosymmetric superconductors }
\author{Keiji Yada$^{1}$, Masatoshi Sato$^{2}$, 
Yukio Tanaka$^{3}$, and Takehito Yokoyama$^{4}$, 
}
\affiliation{
$^1$ Venture Business Laboratory, Nagoya University,Nagoya, 464-8603, Japan \\
$^2$ Institute for Solid State Physics, University of Tokyo, Kashiwanoha 5-1-5, Kashiwa, Chiba 277-8581, Japan\\
$^3$ Department of Applied Physics, Nagoya University,Nagoya, 464-8603, Japan \\
$^4$ Department of Physics, Tokyo Institute of Technology, Tokyo, 152-8551,
Japan 
\\}
\date{\today}

\begin{abstract}
We study Andreev bound state (ABS) and surface density of state (SDOS)
of non-centrosymmetric superconductor where spin-singlet $d$-wave pairing mixes with spin-triplet  $p$ (or $f)$-wave one by spin-orbit coupling.  
For $d_{xy} + p$-wave pairing, ABS appears as a zero energy state.
The present ABS is 
a Majorana edge mode  preserving the time reversal symmetry. 
We calculate topological invariant number and discuss the relevance to 
a single Majorana edge mode. 
In the presence of the Majorana edge mode, the SDOS 
depends strongly on the direction of the 
Zeeman field. 
\end{abstract}

\maketitle



%

%



\section{Introduction}

Recently, physics of non-centrosymmetric (NCS) 
superconductors is one of the important issues in condensed 
matter physics. Actually,  several 
NCS superconductors have been discovered 
such as  CePt$_3$Si \cite{Bauer},  
Li$_{2}$Pt$_{3}$B \cite{Zheng} and LaNiC$_{2}$ \cite{Hillier}.  
Also, the two-dimensional NCS 
superconductivity is expected at the interfaces 
and/or surfaces due to the 
strong potential gradient. 
An interesting example is
the superconductivity at 
LaAlO$_3$/SrTiO$_3$ heterointerface  \cite{Interface}. 
In NCS superconductors, the spin-orbit coupling comes into play. 
One of the remarkable features is that due to the 
broken inversion symmetry, superconducting pair potential becomes a mixture of 
spin-singlet even-parity and spin-triplet odd-parity \cite{Gorkov}. 
Frigeri et al. \cite{Frigeri} have shown that 
$p$ $(p_{x} \pm ip_{y})$-wave pairing 
state has the highest $T_{\rm c}$ within the triplet-channel in CePt$_3$Si. 
Due to the mixture of singlet $s$-wave and triplet $p$-wave 
pairings, several novel properties 
such as the large upper critical field are expected \cite{Frigeri,Fujimoto1}. \par
Up to now, there have been several studies about 
superconducting profiles for NCS superconductors 
\cite{Frigeri,Fujimoto1,Yanase,Iniotakis,Eschrig,Tanaka,Sato2009,Yip}. 
In these work, pairing symmetry of NCS superconductors has been 
mainly assumed to be $s+p$-wave 
where spin-triplet $p (p_{x} \pm i p_{y})$-wave and spin-singlet $s$-wave pair potential 
mixes each other as a bulk state. 
However, in a strongly correlated system, different types of 
pairing symmetries are possible. 
Microscopic calculations have shown that spin-singlet $d_{x^{2}-y^{2}}$-wave 
pairing mixes with spin-triplet $f$-wave pairing based 
on the Hubbard model near half filling \cite{NCS-theory}. 
The magnitude of 
spin-triplet $f$-wave pairing in this $d_{x^{2}-y^{2}} +f$-wave pairing 
is enhanced by Rashba-type spin-orbit coupling originating from the
broken inversion symmetry. 
Also, a possible pairing symmetry of superconductivity generated at heterointerface LaAlO$_3$/SrTiO$_3$ \cite{Interface} has been studied 
based on  a similar model \cite{Yada}. 
In Ref. \cite{Yada}, it has been found that the gap function consists 
of spin-singlet $d_{xy}$-wave component and spin-triplet 
$p$-wave one.
The ratio of the $d_{xy}$-wave and the 
$p_{x}(p_{y})$-wave component 
in this $d_{xy}+p$-wave model continuously changes with the 
carrier concentration.
 \par
Stimulated by these backgrounds, a study of 
Andreev bound states of 
$d_{xy}+p$ or $d_{x^{2}-y^{2}}+f$-wave pairing has started \cite{Mizuno}. 
It has been known that 
the generation of Andreev bound state (ABS) at the surface 
or interface is a remarkable feature  specific to unconventional 
pairing \cite{ABS} 
since ABS directly manifests itself in the tunneling spectroscopy. 
Actually, for $d_{xy}$-wave pairing, zero energy ABS appears \cite{Hu}.
The presence of the ABS 
has been verified by tunneling experiments of high-$T_{\rm c}$ 
cuprate \cite{Hu,TK95} as a zero bias conductance peak \cite{Experiments}. 
For chiral $p$-wave superconducting state realized in 
Sr$_{2}$RuO$_{4}$ \cite{Maeno},  
ABS is generated as a chiral edge mode which has a dispersion proportional to 
the momentum parallel to the interface \cite{Matsumoto}.  
For $s+p$-wave NCS superconductors, 
when the magnitude of $p$-wave pair potential is larger than 
that of $s$-wave one, 
it has been shown that 
ABS is generated at its edge as helical edge modes similar to those in quantum 
spin Hall system\cite{Iniotakis,Eschrig,Tanaka,Kane}. 
Then, several new features of  spin transport
stemming from these helical edge modes 
have been also predicted \cite{Eschrig,Tanaka,Sato2009,Yip}.

In Ref. \cite{Mizuno}, we have clarified the ABS and tunneling conductance $\sigma_{\rm C}$ in normal metal /
NCS superconductor junctions for $d_{xy}+p$-wave and $d_{x^{2}-y^{2}}+ 
f$-wave pairings.
Both for them, 
new types of ABS appear, 
in stark contrast to $s+p$-wave case. 
In particular, for $d_{xy}+p$-wave case, 
due to the existence of the Fermi surface splitting by spin-orbit coupling,
there appears a Majorana edge state with flat dispersion
preserving the time reversal symmetry.
Reflecting the Majorana edge state, $\sigma_{\rm C}$ has a zero bias
conductance peak (ZBCP) in the presence of the spin orbit coupling.

In the present paper, we study the time-reversal invariant Majorana
edge state in details. In particular, we examine the  local density of state
at surface, i.e. surface density of state (SDOS) for NCS superconductors with
$s+p$, $d_{x^{2}-y^{2}} + f$, and  $d_{xy} + p$-wave pair potential,
respectively,
based on the lattice model Hamiltonian.
For $d_{xy} + p$-wave pairing, we confirm that the existence of the special
ABS appears as a zero energy state,
due to the Fermi surface splitting by the spin-orbit coupling.
The present ABS is a single Majorana edge mode preserving the time
reversal symmetry. 

We also study the topological nature of the Majorana edge
mode with flat dispersion.
The ABS found here is topologically stable against a small
deformation of the  Hamiltonian if the deformation preserves the
time-reversal invariance and the translation invariance along the
direction parallel to the edge.
We introduce a topological invariant number ensuring the existence of
the zero energy ABS, and clarify the relevance to the number of 
Majorana edge modes. 
It is revealed that 
the absolute value of the 
topological number equals to the number of the 
Majorana edge modes. 

Topological aspects of edge states 
have been attracting intensive interests in condensed 
matter physics. Especially, it was highlighted by the 
discovery of the quantum Hall system (QHS) showing the 
accurate quantization of the Hall conductance $\sigma_H$ 
which is related to the topological integer \cite{Girvin,Thouless}. 
It is known that chiral edge state is generated at the 
edge of the sample. 
The concept of the QHS has been generalized to 
the time-reversal ($T$) symmetric system, 
i.e., the quantum spin Hall system 
(QSHS) \cite{Kane,Fu}. 
In QSHS, there exist the helical edge modes, i.e.,
the time-reversal pair of right- and left-going 
one-dimensional modes. 
The above edge modes are 
generated from non-trivial nature of bulk Hamiltonian and 
topologically protected. 
Furthermore, recently, to pursue analogous nontrivial edge state 
including Majorana edge mode in superconducting 
system has become a hot issue  \cite{Qi}.  
Although there have been many studies about edge modes in 
topologically non-trivial superconducting systems \cite{Majorana1,Majorana2}, 
the relation between edge modes (ABSs) and the surface 
density of states has not been fully clarified. \par

In the presence of the single Majorana edge mode, the SDOS 
has an anomalous orientational 
dependence on the Zeeman magnetic field. 
We reveal that the SDOS with zero energy peak is robust against 
magnetic field in a certain applied direction.

The organization of the present paper is as follows. 
In Sec. II, we introduce the Hamiltonian and the lattice 
Greenfs function formalism. In Sec. III, the results of the 
numerical calculations of SDOS, topological invariant number, and 
SDOS in the presence of Zeeman magnetic field 
are discussed. 
In Sec. IV, the conclusions and outlook
are presented.

\section{Formulation }
In this paper, we consider the two dimensional square lattice with
Rashba-type spin-orbit coupling.
The model Hamiltonian is given by
\begin{eqnarray}
\mathcal{H}_0&=&\sum_{{\bm k}\sigma}\varepsilon_{\bm k}c^\dag_{{\bm k}\sigma}c_{{\bm k}\sigma}
+\lambda\sum_{\bm k}
{\bm g}({\bm k})\cdot {\bm\sigma}_{\sigma\sigma'}c^\dag_{{\bm
k}\sigma}c_{{\bm k}\sigma'}\nonumber\\ 
&&+\frac{1}{2}\sum_{\bm k}
\{\Delta_{\sigma\sigma'}({\bm k})c^\dag_{{\bm k}\sigma}c^\dag_{-{\bm
k}\sigma'}+{\rm h.c.}\}\nonumber\\
&&-\mu_{\rm B}\sum_{{\bm k}\sigma\sigma'}{\bm H}\cdot{\bm \sigma}_{\sigma\sigma'}c^\dag_{{\bm k}\sigma}c_{{\bm k}\sigma'},
\end{eqnarray}
where $c_{{\bm k}\sigma}$ ($c^\dag_{{\bm k}\sigma}$) is an annihilation
(creation) operator for an electron, $\hat{\bm \sigma}$ and
$\hat\sigma_i$ the Pauli matrices, and $\varepsilon_{\bm k}$  
the energy dispersion of the electron on the square lattice, 
$\varepsilon_{\bm k}=-2t(\cos
k_x+\cos k_y)-\mu$ with the nearest neighbor hopping $t$ and  the chemical
potential $\mu$.
The second term is the Rashba spin-orbit coupling, ${\bm g}({\bm k})=(\sin k_y, -\sin k_x, 0)$,
and the third one is the pair potential,
 $\hat\Delta_{\bm k}=i\psi({\bm k})\hat\sigma_y+i{\bm d}({\bm k})\cdot
 \hat{\bm \sigma}\hat\sigma_y$.
In the presence of the Rashba spin-orbit coupling, the Fermi surfaces
are split into two, and
we suppose intraband pairings in each spin-split bands.
Then the $d$-vector of pairing function for triplet pairings ${\bm
d}({\bm k})$ is aligned with the polarization vector of the Rashba spin-orbit coupling, 
${\bm d}({\bm k})\parallel{\bm g}({\bm k})$.
As a result, 
the triplet component of the energy gap function 
is given by ${\bm d}({\bm k})=\Delta_{\rm t}f({\bm k}){\bm g}({\bm k})$
while that of 
singlet component reads $\psi({\bm k})=\Delta_{\rm s}f({\bm k})$.
Here $f({\bm k})$ is given by $f({\bm k})=1$, $\sin k_x\sin k_y$ and $(\cos
k_x-\cos k_y)$ for $s$+$p$, $d_{xy}$+$p$ and $d_{x^2-y^2}$+$f$-wave,
respectively.\cite{dxy} 
We also introduce the Zeeman splitting term in an applied magnetic field
$\mu_{\rm B}{\bm H}$ for later use.
From the above Hamiltonian, we have the following retarded Green's
function in the infinite system, 
\begin{eqnarray}
\{\check G^{0R}({\bm k},\omega)\}^{-1}&=&(\omega+i\eta)\check I_{4\times4}
-\check {\mathcal{H}}({\bm k}),\label{eq:g}
\end{eqnarray}
with
\begin{eqnarray}
\check {\mathcal{H}}({\bm k})&=&\left(
\begin{array}{cc}
\hat\xi_{\bm k}
&\hat\Delta_{\bm k}\\
\hat\Delta^\dag_{\bm k}&-\hat\xi^*_{-{\bm k}}
\end{array}
\right),
\label{eq:BdG}
\end{eqnarray}
where $\hat\xi_{\bm k}=\varepsilon_{\bm k}\hat I_{2\times2}+\lambda {\bm g}({\bm k})\cdot \hat{\bm \sigma}-\mu_{\rm B}{\bm H}\cdot\hat{\bm \sigma}$ with
the ${4\times 4}$ (${2\times 2}$) unit matrix $\check I_{4\times
4}$($\hat I_{2\times 2}$).

\begin{figure}[htbp]
\begin{center}
\includegraphics[width=6.0cm]{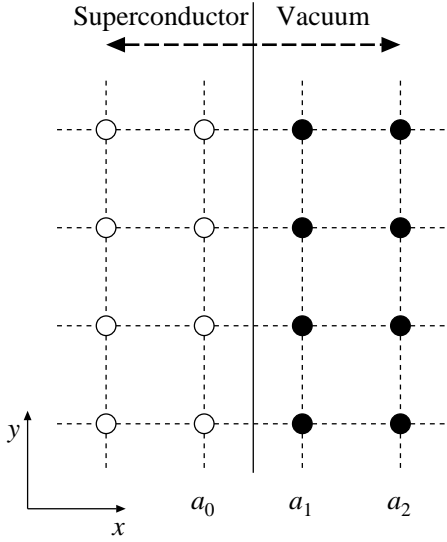}
\caption{(100) surface of square lattice. White and black circles show the sites without and with potential $V$, respectively.}\label{fig1}
\end{center}
\end{figure}

To calculate SDOS, we construct the Green's
function in semi-infinite system.
In the actual numerical calculation, we use the periodic boundary condition
along the $x$-direction with a sufficiently large size of mesh.
To prepare the (100) surface at $x=a_0$, 
we introduce vacuum layers on the right side of the
(100) surface as shown in Fig. \ref{fig1}. 
Here, it is sufficient to introduce the two vacuum layers 
since there is no long range hopping or long range pairing over three
lattice constant in the present model.
On the vacuum layers, we add the following term to the Hamiltonian,
\begin{eqnarray}
\mathcal{H}'= V\sum_{x_i=a_1,a_2}\sum_{\sigma}n_{i\sigma},
\end{eqnarray}
where $n_{i\sigma}=c^\dag_{i\sigma}c_{i\sigma}$ is a number operator at
site $i$ with spin $\sigma$, and $V$ is the on-site potential.
In the limit $V\rightarrow\infty$, no electron exists on the vacuum layers.

First, we switch on the potential $V$ only at the site $x=a_1$.
The Green function $\check G^{1R}(x_1,x_2;k_y,\omega)$ in this situation
satisfies the following equation,
\begin{eqnarray}
\check G^{1R}(x_1,x_2;k_y,\omega)&=&\check G^{0R}(x_1,x_2;k_y,\omega)\nonumber\\
&&\hspace{-3cm}+\check G^{0R}(x_1,a_1;k_y,\omega)V\check\tau_3\check
 G^{1R}(a_1,x_2;k_y,\omega),
\label{eq4}
\end{eqnarray}
where
$\check G^{0R}(x_1,x_2;k_y,\omega)$ is the Fourier component 
$\check G^{0R}({\bm k},\omega)$ with respect to $k_x$,
\begin{eqnarray}
\check G^{0R}(x_1,x_2;k_y,\omega)&=&\frac{1}{N_x}\sum_{k_x}
\check G^{0R}(\bm{k},\omega)e^{ik_x(x_1-x_2)}.
\end{eqnarray}
Here $N_x$ is the number of $x$-meshes, and $\check\tau_3$ is the Pauli
matrix in the particle-hole space. 
\begin{eqnarray}
\check\tau_3=
\left(
\begin{array}{cc}
\hat I_{2\times2} & 0\\
0 & -\hat I_{2\times2}
\end{array}
\right).
\end{eqnarray}
In the  $V\rightarrow\infty$ limit,
we have the following solution of Eq. (\ref{eq4}),
\begin{eqnarray}
\check G^{1R}(x_1,x_2;k_y,\omega)&=&\check G^{0R}(x_1,x_2;k_y,\omega)\nonumber\\
&&\hspace{-1cm}-\check G^{0R}(x_1,a_1;k_y,\omega)\{\check G^{0R}(a_1,a_1;k_y,\omega)\}^{-1}\nonumber\\
&&\hspace{1.5cm}\times\check G^{0R}(a_1,x_2;k_y,\omega).
\label{eq:g1r}
\end{eqnarray}

Then, we switch on the potential $V$ at the site $x=a_2$ as well.
For the Green's function $\check G^{2R}(x_1,x_2;k_y,\omega)$ in this
case, 
we have the following equation,
\begin{eqnarray}
\check G^{2R}(x_1,x_2;k_y,\omega)&=&\check G^{1R}(x_1,x_2;k_y,\omega)\nonumber\\
&&\hspace{-3cm}+\check G^{1R}(x_1,a_2;k_y,\omega)V\check\tau_3\check
 G^{1R}(a_2,x_2;k_y,\omega),
\end{eqnarray}
Therefore, taking the $V\rightarrow\infty$ limit, we obtain the Green's
function in the semi-infinite system,
\begin{eqnarray}
\check G^{2R}(x_1,x_2;k_y,\omega)&=&\check G^{1R}(x_1,x_2;k_y,\omega)\nonumber\\
&&\hspace{-1cm}-\check G^{1R}(x_1,a_2;k_y,\omega)\{\check G^{1R}(a_2,a_2;k_y,\omega)\}^{-1}\nonumber\\
&&\hspace{1.5cm}\times\check G^{1R}(a_2,x_2;k_y,\omega).
\label{eq:g2r}
\end{eqnarray}
From (\ref{eq:g1r}) and (\ref{eq:g2r}), we can calculate the SDOS $\rho_{\rm s}(\omega)$, which is given by the
local density of states at $x=a_0$,
\begin{eqnarray}
\rho_{\rm s}(\omega) =-\frac{1}{N_y}\sum_{k_y}\sum_{\alpha=1,2}{\rm Im}\{G^{2R}_{\alpha\alpha}(a_0,a_0;k_y,\omega)\},
\end{eqnarray}
while the DOS in the bulk is given by
\begin{eqnarray}
\rho_{\rm b}(\omega)=-\frac{1}{N_xN_y}\sum_{k_xkS_y}\sum_{\alpha=1,2}{\rm Im}\{G^{0R}_{\alpha\alpha}({\bm k},\omega)\},
\end{eqnarray}
where $N_x$ and $N_y$ are the number of meshes along the $x$ and the $y$ direction, respectively.
In the calculation presented in the following, we choose $N_x=N_y=2^{13}$. 
We set $t=\Delta_0=1$ for the unit of energy.
The number of electrons per unit cell is $0.3$ ($\mu\sim-2.40$).
To guarantee the convergence, we use $\eta=0.03t$ for the
infinitesimal imaginary parts in Eq. (\ref{eq:g}).

\section{Results}
In this section, we first focus on the local density of state at the surface, 
$i.e.$, SDOS $\rho_{\rm s}(\omega)$ 
for various paring symmetry. 
Since spin-singlet and spin-triplet components of pair potential 
mix in general, we introduce a parameter $r_{s}$, 
which denotes the ratio of singlet component. 
The parameter $r_{s}$ ($0\ge r_{s}\ge1$) is defined as 
$\Delta_{s}=r_{s}\Delta_{0}$ and $\Delta_{t}=(1-r_{s})\Delta_{0}$.

\begin{figure}[htbp]
\begin{center}
\includegraphics[width=7.5cm]{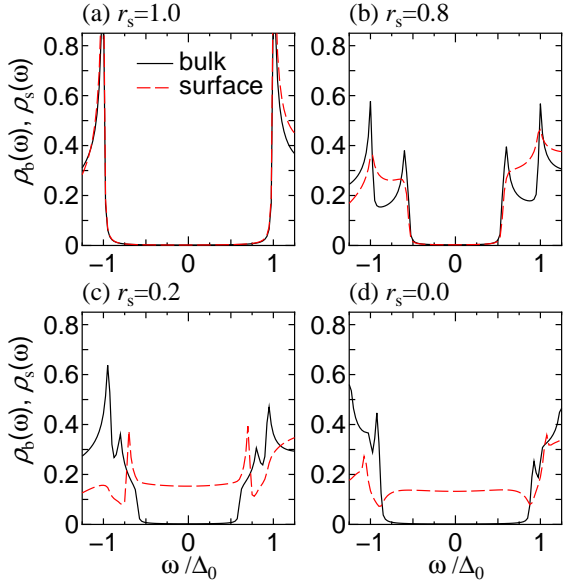}
\caption{(Color online) Local density of states for $s$+$p$-wave in the bulk (solid lines) and at the surface (dashed lines) for $\lambda=0.5$ and $r_s=1.0$, $0.8$, $0.2$, and $0.0$.}\label{fig2}
\end{center}
\end{figure}

\begin{figure}[htbp]
\begin{center}
\includegraphics[width=8.5cm]{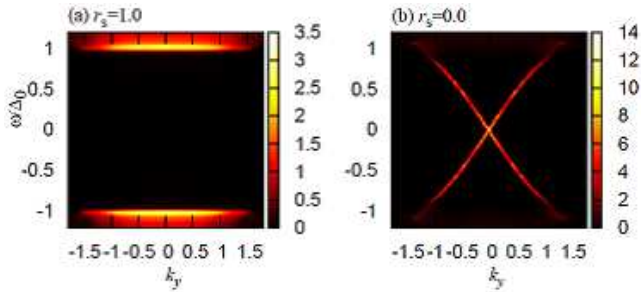}
\caption{(Color online) Angle resolved local density of state 
of $s$+$p$-wave pairing is plotted 
as a function of $k_{y}$ with $\lambda=0.5$.
(a)$\Delta_{s}=\Delta_{0}$ and $\Delta_{t}=0$,
(b)$\Delta_{t}=\Delta_{0}$ and $\Delta_{s}=0$.}
\label{fig3}
\end{center}
\end{figure}

The SDOS for $s+p$-wave case is plotted in Fig. \ref{fig2}. 
In this case, the bulk DOS $\rho_{\rm b}(\omega)$ 
always has a $U$-shaped gap structure 
as shown in solid lines. The magnitude of the gap is given by 
$\mid \Delta_{s}-\Delta_{t} \mid$. 
For $\Delta_{s}>\Delta_{t}$,
the SDOS $\rho_{\rm s}(\omega)$ also has a $U$-shaped gap structure
as shown in the dashes lines in Fig. \ref{fig2} (a) and (b).
On the other hand, for $\Delta_{t}>\Delta_{s}$ with $r_{s}=0.2$ and
$r_{s}=0$, 
the resulting 
$\rho_{\rm s}(\omega)$ has a residual value at $\omega=0$ 
as shown in Figs.\ref{fig2}(c) and (d), respectively.
To show up the SDOS more clearly,
we also show the angle resolved surface density of states (ARSDOS)
$-\sum_{\alpha=1,2}{\rm Im}\{G^{2R}_{\alpha\alpha}(a_0,a_0;k_y,\omega)\}$.
As shown in Fig. \ref{fig3}(a), ARSDOS shows a full gap structure without any inner gap state
for singlet dominant case ($\Delta_{s}>\Delta_{t}$).
On the other hand, as shown in Fig. \ref{fig3}(b),
the ARSDOS for triplet dominant case ($\Delta_{t}>\Delta_{s}$) has two branches of Andreev bound state,
which are dubbed as helical edge modes \cite{Tanaka,Sato2009}.
The presence of helical edge modes inside the bulk energy gap
induces the residual SDOS inside the bulk energy gap.


\begin{figure}[htbp]
\begin{center}
\includegraphics[width=7.5cm]{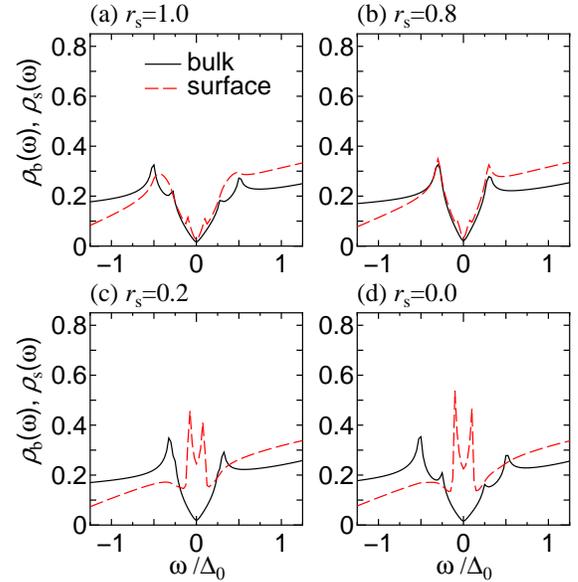}
\caption{(Color online) Local density of states for $d_{x^2-y^2}$+$f$-wave in the bulk (solid lines) and at the surface (dashed lines) for $\lambda=0.5$ and $r_s=1.0$, $0.8$, $0.2$, and $0.0$.}\label{fig4}
\end{center}
\end{figure}

\begin{figure}[htbp]
\begin{center}
\includegraphics[width=8.5cm]{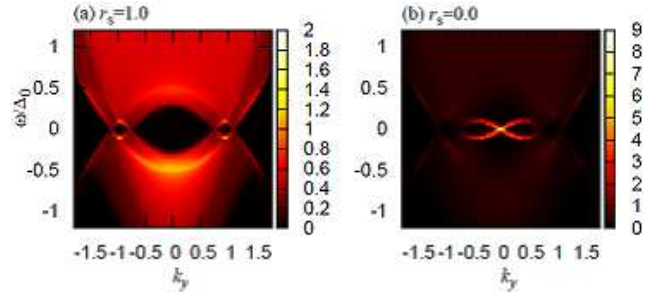}
\caption{(Color online) Angle resolved local density of state 
of $d_{x^2-y^2}$+$f$-wave pairing is plotted 
as a function of $k_{y}$ with $\lambda=0.5$.
(a)$\Delta_{s}=\Delta_{0}$ and $\Delta_{t}=0$,
(b)$\Delta_{t}=\Delta_{0}$ and $\Delta_{s}=0$.}
\label{fig5}
\end{center}
\end{figure}

Next we look at SDOS for $d_{x^{2}-y^{2}}+f$-wave case plotted in Fig. \ref{fig4}. 
In this case, $\rho_{\rm b}(\omega)$ 
 always has a $V$-shaped gap structure as shown in solid lines 
reflecting the nodal structures of the bulk energy gap.  
The corresponding $\rho_{\rm s}(\omega)$ (dashed line) 
also has a similar $V$-shaped gap structure 
for singlet dominant case ($\Delta_{s}>\Delta_{t}$) 
with $r_{s}=1$ and $r_{s}=0.8$. 
As shown in Fig.\ref{fig5}(a), there is no inner gap state in ARSDOS,
while the bulk energy gap has a strong $k_{y}$-dependence. 
On the other hand, for triplet dominant case ($\Delta_{t}>\Delta_{s}$) 
with $r_{s}=0.2$ and $r_{s}=0$, 
$\rho_{\rm s}(\omega)$ has two additional peaks 
as shown in dashed lines of 
Figs.\ref{fig4}(c) and (d) as compared to the bulk density of states 
\cite{Mizuno} (solid lines). 
The additional two peaks in the SDOS
originates from the helical edge modes 
generated inside energy gap as shown in 
Fig. \ref{fig5}(b) \cite{Mizuno}. \par


\begin{figure}[htbp]
\begin{center}
\includegraphics[width=7.5cm]{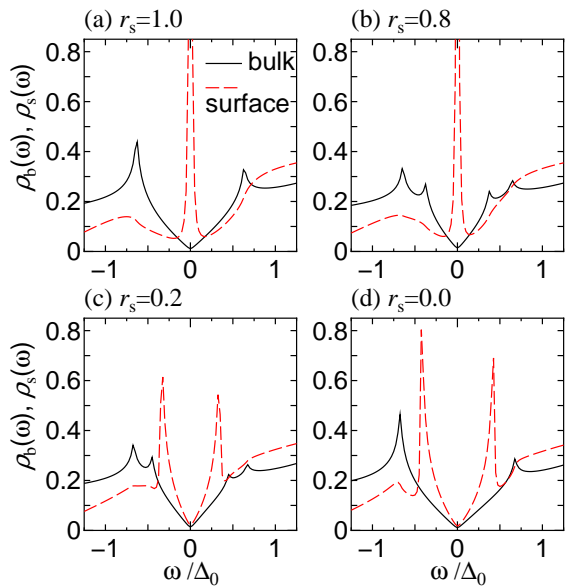}
\caption{(Color online) Local density of states for $d_{xy}$+$p$-wave in the bulk (solid lines) and at the surface (dashed lines) for $\lambda=0.0$ and $r_s=1.0$, $0.8$, $0.2$, and $0.0$.}\label{fig6}
\end{center}
\end{figure}
\begin{figure}[htbp]
\begin{center}
\includegraphics[width=8.5cm]{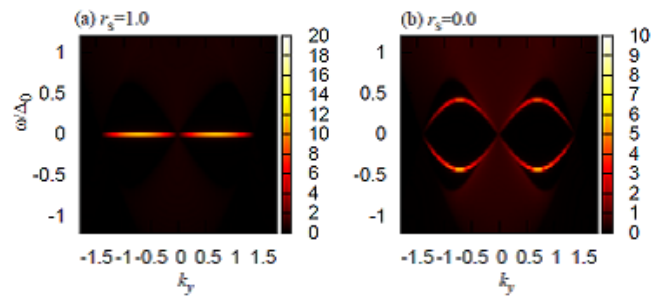}
\caption{(Color online) Angle resolved local density of state 
of $d_{xy}$+$p$-wave pairing is plotted 
as a function of $k_{y}$ with $\lambda=0.0$.
(a)$\Delta_{s}=\Delta_{0}$ and $\Delta_{t}=0$,
(b)$\Delta_{t}=\Delta_{0}$ and $\Delta_{s}=0$.}
\label{fig7}
\end{center}
\end{figure}
\begin{figure}[htbp]
\begin{center}
\includegraphics[width=7.5cm]{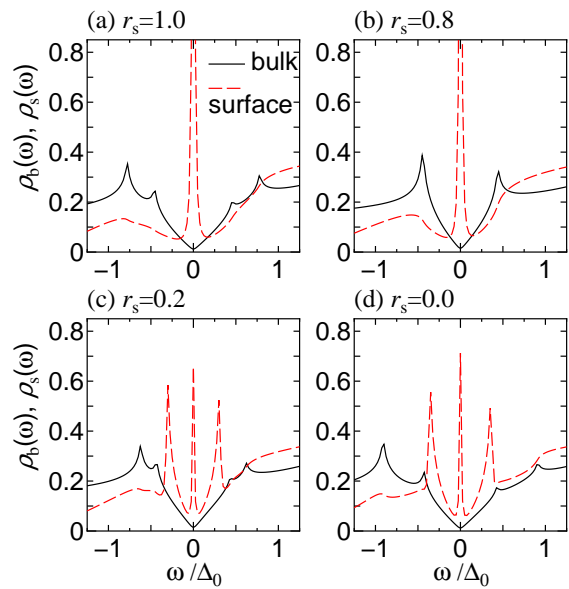}
\caption{(Color online) Local density of states for $d_{xy}$+$p$-wave in the bulk (solid lines) and at the surface (dashed lines) for $\lambda=0.5$ and $r_s=1.0$, $0.8$, $0.2$, and $0.0$.}\label{fig8}
\end{center}
\end{figure}

The SDOS for $d_{xy}+p$-wave NCS superconductors is plotted in 
Figs. \ref{fig6} and \ref{fig8}. 
In this case, $\rho_{\rm s}(\omega)$ 
has very different line shapes as compared to former two cases. 
In particular, for the spin-triplet dominant pairing,  we find that the
SDOS is very sensitive to the spin-orbit coupling. 
To show this, we first start with the case 
without spin-orbit coupling {\it i.e.} $\lambda=0$. See Fig. \ref{fig6}. 
$\rho_{\rm b}(\omega)$ has a $V$-shaped gap structure 
as shown in solid lines, reflecting on the nodal structure of pair potential.  
The corresponding $\rho_{\rm s}(\omega)$ (dashed line) 
has a  zero energy peak (ZEP) for singlet dominant case  with $r_{s}=1$ and $r_{s}=0.8$. 
This ZEP comes from the mid gap Andreev bound state
with flat dispersion as shown in Fig. \ref{fig7}(a),
and it is essentially the same as that appears in the surface state of high-$T_{\rm c}$ cuprate \cite{Hu,TK95}.
On the other hand, for triplet dominant case with 
$r_{s}=0.2$ and $r_{s}=0$, the ZEP disappear, but  
$\rho_{\rm s}(\omega)$ supports two additional peaks 
instead, as shown in dashed lines of 
Figs.\ref{fig6}(c) and (d), respectively. 
The additional two peaks are generated by  
the anomalous ABS.
In the absence of Rashba spin-orbit coupling,
the dispersion of the anomalous ABS for $d_{xy}+p$-wave pairing is very similar
to the helical edge modes for $d_{x^2-y^2}+f$-wave pairing,
as shown in Fig. \ref{fig7}(b). 
However, the intensity of ARSDOS near $k_y\sim0$ is very low
since the magnitude of the excitation energy of the ABS is close to the bulk energy gap.
In the 2D free electron model for NCS superconductor,
it is shown that the dispersion of ABS corresponds the bulk energy gap
and the intensity of ARSDOS is completely absent for $|k_{y}|<k_c$.\cite{Mizuno}.
Thus, in contrast to the helical edge mode in the $d_{x^2-y^2}+f$-wave pairing case,
the values of the SDOS $d_{xy}+p$-wave pairing at $\omega/\Delta=0$ is very close to zero.


Let us now consider the $d_{xy}+p$-wave case with nonzero spin-orbit coupling 
$\lambda$. As shown in Fig. \ref{fig8}, the bulk DOS 
$\rho_{\rm b}(\omega)$ shows a $V$-shaped gap structure similar to 
Fig.\ref{fig6}. 
Then, 
for singlet dominant case  with
$r_{s}=1$ and $r_{s}=0.8$, the SDOS  $\rho_{s}(\omega)$ (dashed line in
Figs. \ref{fig8}(a) and (b), respectively)  
has a ZEP 
similar to Figs. \ref{fig6}(a) and (b). 
On the other hand, for triplet dominant case with 
$r_{s}=0.2$ and $r_{s}=0$, in addition to two peaks similar to those in 
Figs. \ref{fig6}(c) and \ref{fig6}(d), ZEP appears as shown in Figs.\ref{fig8}(c) and \ref{fig8}(d). 
It is remarkable that ZEP is generated by spin-orbit coupling $\lambda$ 
\cite{Mizuno}.

\begin{figure}[htbp]
\begin{center}
\includegraphics[width=8.5cm]{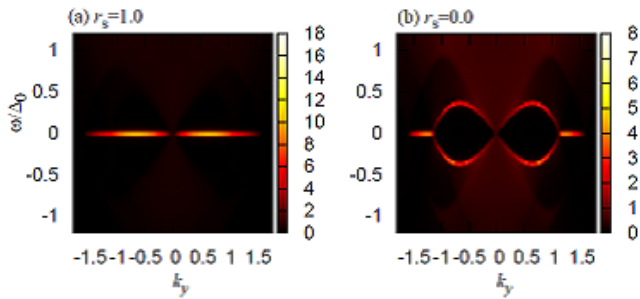}
\caption{(Color online) Angle resolved local density of state 
of $d_{xy}$+$p$-wave pairing is plotted 
as a function of $k_{y}$ with $\lambda=0.5$.
(a)$\Delta_{s}=\Delta_{0}$ and $\Delta_{t}=0$,
(b)$\Delta_{t}=\Delta_{0}$ and $\Delta_{s}=0$.}
\label{fig9}
\end{center}
\end{figure}

To show up the new ZEP much more clearly, we plot ARSDOS of
$d_{xy}+p$-wave case in Fig. \ref{fig9} with $\lambda=0.5$. 
Comparing with the ARSDOS in Fig.\ref{fig7}, for triplet dominant case, 
we find that an additional zero energy state (ZES)
appears.
In the presence of the spin-orbit coupling, the Fermi surface is
split into the large one with the Fermi momenta $k_2$ and the small one with
$k_1$.
The ZES exists only for $k_y$ between the split Fermi surfaces, namely,
for $k_y$ with $k_2>|k_y|>k_1$. 
It can be shown that the Bogoliubov quasiparticle creation
operator $\gamma_{\bm k}^{\dagger}$ for the ZES satisfies $\gamma_{\bm
k}^{\dagger}=\gamma_{-{\bm k}}$.\cite{Mizuno}
Therefore, the ZES is identified as a Majorana Fermion.
  
For singlet dominant $d_{xy}+p$-wave NCS superconductors,  
we obtain the SDOS and ARSDOS depicted in Figs.\ref{fig8}(a)-(b) and 
Fig.\ref{fig9}(a), respectively. 
At first sight, they look very similar to those in
Figs.\ref{fig7}(a)-(b) and Fig.\ref{fig8}.  
However, for $k_y$ with $k_2>|k_y|>k_1$, we again have a single branch of a
zero energy state on the edge. 
As is shown later, because of the existence of
the time-reversal invariant Majorana Fermion (TRIMF), the SDOS
for the $d_{xy}+p$-wave NCS superconductor shows a peculiar dependence on the
Zeeman magnetic field.

Unlike the Majorana Fermions studied before, the present Majorana
Fermion is realized with the time reversal invariance.
The TRIMF has the following
three characteristics. 
(a) It has a unique flat dispersion. To be consistent with the
time-reversal invariance, the single branch of ZES should be symmetric
under $k_y\rightarrow -k_y$. Therefore, by taking into account the
particle-hole symmetry as well, the flat dispersion is required.
On the other hand, the conventional time-reversal breaking Majorana
edge state has a linear dispersion.
(b) The spin-orbit coupling is indispensable for the existence of
the TRIMF.
Without the spin-orbit coupling, the TRIMF vanishes. 
(c) The TRIMF is topologically stable under small deformation of the
Hamiltonian. 
The topological stability is ensured by a
topological number, which will be shown below.

%
%

To see the topological nature of the TRIMF, 
we start from the Bogoliubov-de Gennes (BdG) Hamiltonian in the bulk
system defined in eq. (\ref{eq:BdG}) without Zeeman magnetic field.
In the Nambu representation, the BdG Hamiltonian (\ref{eq:BdG}) has the
particle-hole symmetry, 
\begin{eqnarray}
\check C \check {\mathcal{H}}({\bm k})\check C^{\dagger}=-\check {\mathcal{H}}^{*}(-{\bm k}),
\quad
\check C=\left(
\begin{array}{cc}
0&\hat I_{2\times2}\\
\hat I_{2\times2}&0
\end{array}
\right).
\end{eqnarray}
In addition, from the time-reversal invariance, the BdG Hamiltonian
satisfies 
\begin{eqnarray}
\check\Theta \check {\mathcal{H}}({\bm k}) \check\Theta ^{\dagger}=\check {\mathcal{H}}^{*}(-{\bm k}), 
\quad
\check\Theta=\left(
\begin{array}{cc}
i\sigma_y&0\\
0&i\sigma_y
\end{array}
\right).
\end{eqnarray}
Therefore, we can define the operator $\Gamma$ which anticommutes with
the BdG Hamiltonian,
\begin{eqnarray}
\{\check {\mathcal{H}}({\bm k}), \check\Gamma\}_+=0. 
\end{eqnarray}
Here $\Gamma$ is defined as the product of 
particle-hole transformation operator $C$ and time reversal operator $\Theta$,
\begin{eqnarray}
\check\Gamma=-i\check C\check\Theta=\left(
\begin{array}{cc}
0&\hat\sigma_y\\
\hat\sigma_y&0
\end{array}
\right).
\end{eqnarray}
Now we take the basis which diagonalizes $\check\Gamma$  
\begin{eqnarray}
\check U_\Gamma^\dag \check\Gamma \check U_\Gamma=\left(
\begin{array}{cc}
\hat I_{2\times2}&0\\
0&-\hat I_{2\times2}
\end{array}
\right)
\end{eqnarray}
with the unitary matrix $\check U_\Gamma$
\begin{eqnarray}
\check U_\Gamma=\check U_\Gamma^\dag=\frac{1}{\sqrt2}\left(
\begin{array}{cc}
\hat I_{2\times2}&\hat \sigma_y\\
\hat \sigma_y&-\hat I_{2\times2}
\end{array}
\right).
\end{eqnarray}
Then, we find that the BdG Hamiltonian $\check {\mathcal{H}}({\bm k})$ becomes
off-diagonal in this basis,
\begin{eqnarray}
\check U_\Gamma^\dag\check {\mathcal{H}}({\bm k})\check U_\Gamma=\left(
\begin{array}{cc}
0&\hat q({\bm k})\\
\hat q({\bm k})^\dag&0
\end{array}
\right),
\end{eqnarray}
where $\hat q({\bm k})=\hat\xi_{\bm k}\hat\sigma_y-\hat\Delta_{\bm k}$.

To classify the existence or nonexistence of the ZES with $k_y=k_y^0$,
we change $k_{x}$ from $-\pi$ to $\pi$ 
for a fixed value of $k_y=k_y^0$.
The winding number $W$ is defined as a number of revolutions of
det$|\hat q({\bm k})|\equiv m_1({\bm k})+im_2({\bm k})$  around the origin of
complex plane when $k_{x}$ changes from $-\pi$ to $\pi$, 
\begin{eqnarray}
W(k_y^0)=\frac{1}{2\pi}\int_{-\pi}^{\pi}\left.\frac{\partial\theta({\bm k})}{\partial k_x}\right|_{k_y\rightarrow k_y^0} dk_x, 
\end{eqnarray}
with $\theta({\bm k})\equiv\arg\det|\hat q({\bm k})|=\tan^{-1}(m_2({\bm k})/m_1({\bm k}))$.
The resulting $W(k_y^0)$ must be an integer since the starting point and the
end point of integration route are equivalent in the Brillouin zone.
Therefore, the value of $W(k_y^0)$ changes discretely with the change of $k_y$-value.
When the gap of the system closes at certain points on the integration route,
$W(k_y^0)$ is ill-defined since
det$|\hat q({\bm k})|$ becomes to zero on such points.
Thus, the value of $W(k_y)$ does not change by the small change of $k_y$-value unless the energy gap closes.

\begin{figure}[htbp]
\begin{center}
\includegraphics[width=4.5cm]{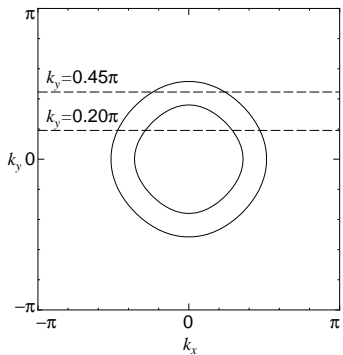}
\caption{Fermi surfaces of spin-split bands due to the Rashba spin-orbit coupling (solid lines) and cutting lines $k_y=0.20\pi$ and $0.45\pi$ (dashed lines).}\label{fig10}
\end{center}
\end{figure}

Here, we study the winding number for the $d_{xy}+p$-wave NCS superconductors
for on the two trajectories for $k_y=0.20\pi$ and $0.45\pi$ as shown in Fig. \ref{fig10}.
For $k_{y}=0.2\pi$, the trajectory crosses the Fermi surfaces four times.
On the other hand, for  $k_{y}=0.45\pi$, it crosses two times.  
As was explained above, the winding number $W(k_y)$ does not change
unless the energy gap closes. 
For $d_{xy}+p$-wave parings, the energy gap closes at $(\pm k_1,0)$, $(\pm k_2,0)$, $(0, \pm k_1)$ and $(0, \pm k_2)$.
Therefore, $W(k_y)$ can change only at $k_y=\pm k_1$, $k_y=\pm k_2$ or $k_y=0$.
Thus, it is sufficient to study the representative point in each $k_y$ range.

\begin{figure}[htbp]
\begin{center}
\includegraphics[width=7.5cm]{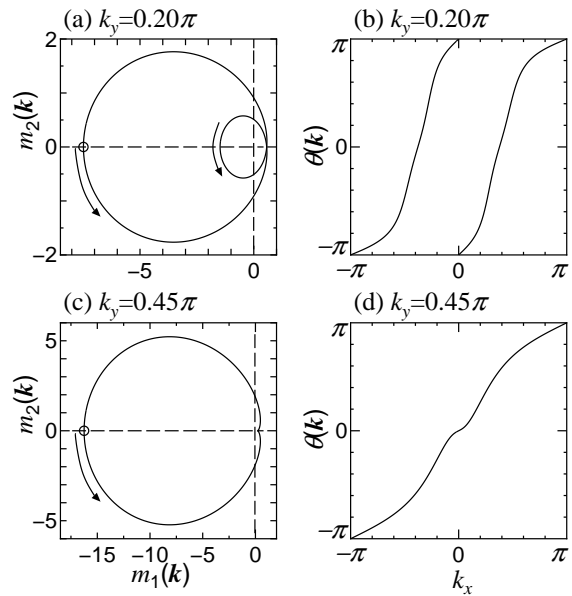}
\caption{Polar plot of $\theta({\bm k})$ for $\Delta_{s}=\Delta_{0}$ and $\Delta_{t}=0$
as a function of $k_{x}$ with fixed $k_{y}$,  
(a) $k_{y}=0.2\pi$ and (c) $k_{y}=0.45\pi$, 
where white circle corresponds to $k_x=\pm\pi$ and arrows show the direction in which the values of $k_x$ increase.
The corresponding 
$\theta=\tan^{-1}(m_{2}/m_{1})$ is plotted as a function of $k_{x}$ 
for (b) $k_{y}=0.2\pi$ and 
(d) $k_{y}=0.45\pi$.}\label{fig11}
\end{center}
\end{figure}

%
We first focus on the singlet dominant $d_{xy}+p$-wave NCS superconductor with $\Delta_{s}=\Delta_{0}$ and 
$\Delta_{t}=0$.
For $k_{y}=0.2\pi$,
as shown in Fig. \ref{fig11}(a), $m_{1}$ and $m_{2}$ draw a curve 
which turns anti-clockwise twice around the $(m_{1},m_{2})=(0,0)$
if we change $k_{x}$ from $-\pi$ to $\pi$. 
At the same time, $\theta({\bm k})$ changes twice 
from $-\pi$ to $\pi$ as shown in Fig. \ref{fig11}(b).  
Therefore, the resulting 
winding number $W$ is $W=2$. 
On the other hand, for $k_{y}=0.45\pi$,  $m_{1}$ and $m_{2}$ draw a curve 
which turns anti-clockwise once around the $(m_{1},m_{2})=(0,0)$ as 
shown in Fig. \ref{fig11}(c).
Moreover, $\theta({\bm k})$ changes once 
from $-\pi$ to $\pi$ as shown in Fig. \ref{fig11}(d).  
Thus the resulting winding number $W$ equals to $W=1$.
This means that these two cases belong to different topological classes. 
In a similar manner, it is possible to generalize above argument for
other $k_{y}$s  ($-\pi < k_{y} < \pi$).
%
We summarize the obtained results in Table \ref{table} (a).

\begin{table*}
\begin{tabular}{c|c||c|c|c|c|c|c|}
\cline{2-8}
&$k_y$ & $k_{y}>k_{2}$ & $k_{2}>k_{y}>k_{1}$ & $k_{1}>k_{y}>0$ & $0>k_{y}>-k_{1}$ & $-k_{1}>k_{y}>-k_{2}$ & $-k_{2}>k_{y}$\\
\cline{2-8}
(a)&$W(k_y)$ & $0$ & $1$ & $2$ & $-2$ & $-1$ & $0$ \\
\cline{2-8}
(b)&$W(k_y)$ & $0$ & $-1$ & $0$ & $0$ & $1$ & $0$ \\
\cline{2-8}
\end{tabular}
\caption{The winding number $W(k_y)$ for $d_{xy}+p$-wave with
(a) $\Delta_{\rm s}>\Delta_{\rm t}$,
(b) $\Delta_{\rm t}<\Delta_{\rm s}$.}\label{table}
\label{table}
\end{table*}

\begin{figure}[htbp]
\begin{center}
\includegraphics[width=7.5cm]{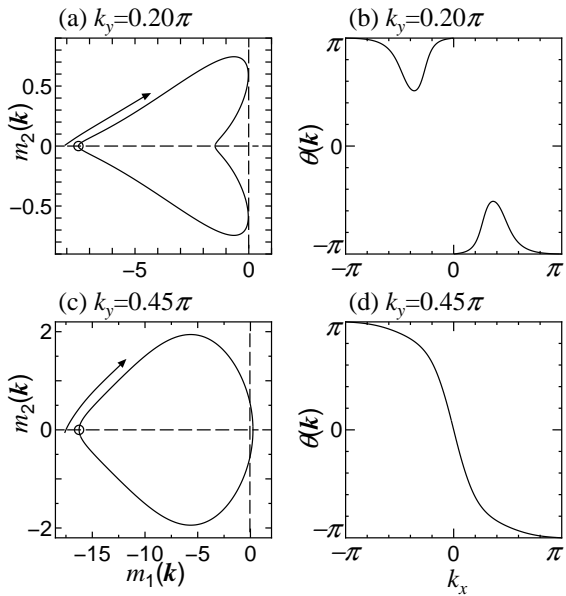}
\caption{Polar plot of $\theta({\bm k})$ for $\Delta_{t}=\Delta_{0}$  and $\Delta_{s}=0$
as a function of $k_{x}$ with fixed $k_{y}$,
(a) $k_{y}=0.2\pi$ and (c) $k_{y}=0.45\pi$,
where white circle corresponds to $k_x=\pm\pi$ and arrows show the direction in which the values of $k_x$ increase.
The corresponding 
$\theta=\tan^{-1}(m_{2}/m_{1})$ is plotted as a function of $k_{x}$ 
for (b) $k_{y}=0.2\pi$ and 
(d) $k_{y}=0.45\pi$.}\label{fig12}
\end{center}
\end{figure}

The same plot of $m_i({\bm k})$ and $\theta({\bm k})$ for the triplet dominant $d_{xy}+p$-wave NCS
superconductors with $\Delta_{t}=\Delta_{0}$ and 
$\Delta_{s}=0$ is shown in Fig. \ref{fig12}. 
For $k_{y}=0.2\pi$, 
$m_{1}$ and $m_{2}$ draw a curve as shown in Fig. \ref{fig12}(a). 
However, it does not  turn around the $(m_{1},m_{2})=(0,0)$ 
with the change of $k_{x}$ from $-\pi$ to $\pi$. 
At the same time, $\theta({\bm k})$ does not 
change from $-\pi$ to $\pi$ as shown in Fig. \ref{fig12}(b), in contrast to 
that in Fig. \ref{fig11}(b).   
This corresponds to the fact that the resulting 
winding number $W$  equals to $W=0$. 
On the other hand, for $k_{y}=0.45\pi$,  $m_{1}$ and $m_{2}$ draw a curve 
which turns clockwise once around the $(m_{1},m_{2})=(0,0)$ as 
shown in Fig. \ref{fig12}(c). 
Also, $\theta({\bm k})$ changes once 
from $\pi$ to $-\pi$ as shown in Fig. \ref{fig12}(d).  
Therefore, the corresponding winding number $W$  equals to $W=-1$.
In a manner to the singlet dominant case, it is possible to generalize this argument for other $k_{y}$s. 
($-\pi < k_{y} < \pi$)
The obtained results are summarized in Table \ref{table}(b).

Comparing the obtained winding number $W(k_y)$ in Table \ref{table} with the
ARSDOS in Fig \ref{fig9}, we notice that zero energy ABS appears
only for $k_y$ with nonzero $W(k_y)$. 
This correspondence implies that the existence and the stability of the
zero energy ABS is ensured by the winding number.
Indeed, we can say the absolute value of $W$, $i.e.$, 
$\mid W \mid$ equals  to the  number of branch of zero energy ABS
\cite{STYY10}. 
In other words,  the number of TRIMF 
equals to the absolute value  of $W$. 
Furthermore, the TRIMF found here are topologically stable against a small
deformation of the BdG Hamiltonian if the deformation preserves the
time-reversal invariance and the translation invariance along the
direction parallel to the edge, both of which are necessary to define the
winding number.
It should be remarked that both for the singlet dominant case and the
triplet dominant one, a single 
TRIMS is generated for 
$k_{2}>\mid k_{y} \mid>k_{1}$.  \par

As is shown above, 
the time-reversal invariance is essential to define the winding number $W$
ensuring the topological stability of the TRIMF.
Therefore, in general, if we apply a perturbation
breaking the time-reversal invariance, the winding number $W$ becomes
the meaningless. 
In other words, a time-reversal breaking perturbation may change the
SDOS of the TRIMF substantially.

\begin{figure}[htbp]
\begin{center}
\includegraphics[width=7.5cm]{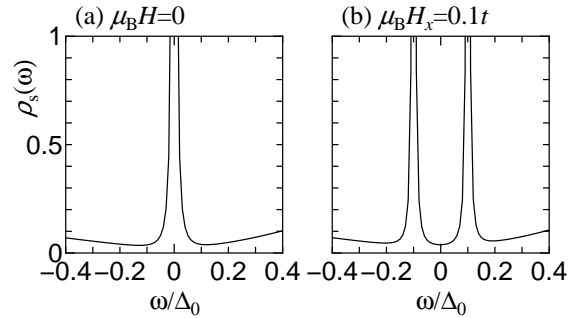}
\caption{ Local density of states at the surface  for $d_{xy}$+$p$-wave
with $\Delta_{s}=\Delta_{0}$ and $\Delta_{t}=0$
for $\lambda=0$ and
(a) $\mu_{\rm B}{\bm H}=0$,
(b) $\mu_{\rm B}H_x=0.1$ (${\bm H}\parallel{\bm x}$).}\label{fig13}
\end{center}
\end{figure}

As a time-reversal breaking perturbation, we consider the Zeeman
magnetic field.
Let us look at the SDOS in the presence of 
Zeeman magnetic field ${\bm H}$.
First, we consider the singlet dominant $d_{xy}+p$-wave NCS superconductor.
Without the spin orbit coupling, the Andreev bound state becomes 
conventional and can be expressed by double TRIMFs. 
In this case, as we expected, it is found that 
the zero energy peak of SDOS from the double TRIMF is split into two by 
any Zeeman magnetic field ${\bm H}$ as shown in
Fig. \ref{fig13} \cite{Kashiwaya98}. 
It is also noted that the resulting SDOS is 
independent of the direction of ${\bm H}$ due to the 
spin-rotational symmetry in the system. 
Thus, the SDOSs $\rho_{s}(\omega)$ for 
$\mu_{B}H_{x}=0.1t$ (${\bm H}\parallel{\bm x}$),
$\mu_{B}H_{y}=0.1t$ (${\bm H}\parallel{\bm y}$) and
$\mu_{B}H_{z}=0.1t$ (${\bm H}\parallel{\bm z}$)
are the same.

\begin{figure}[htbp]
\begin{center}
\includegraphics[width=7.5cm]{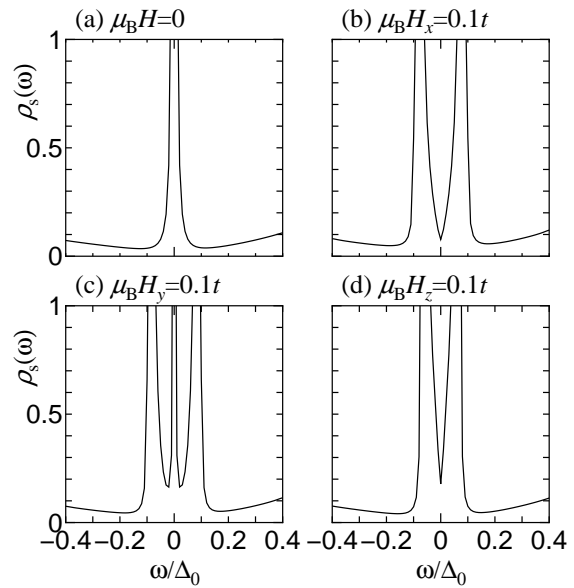}
\caption{  Local density of states at the surface  for $d_{xy}+p$-wave with $\Delta_{s}=\Delta_{0}$ and 
$\Delta_{t}=0$ 
for $\lambda=0.5$ and
(a) $\mu_{\rm B}{\bm H}=0$,
(b) $\mu_{\rm B}H_x=0.1t$ (${\bm H}\parallel{\bm x}$),
(c) $\mu_{\rm B}H_y=0.1t$ (${\bm H}\parallel{\bm y}$),
(d) $\mu_{\rm B}H_z=0.1t$ (${\bm H}\parallel{\bm z}$). 
}\label{fig14}
\end{center}
\end{figure}

On the other hand, in the presence of the spin-orbit coupling, this
property is not satisfied anymore. 
As shown in Fig.\ref{fig14}, $\rho_{s}(\omega)$ has a different values 
for $\mu_{B}H_{x}=0.1t$ [Fig. \ref{fig14}(b)], $\mu_{B}H_{y}=0.1t$ [Fig. \ref{fig14}(c)]
and  $\mu_{B}H_{z}=0.1t$ [Fig. \ref{fig14}(d)], respectively. 
The orientational dependence of SDOS is due to the presence of 
the spin-orbit coupling since the spin-rotational symmetry is broken.  
It is noted, three peak structure including zero energy peak 
appears for $\mu_{\rm B}H_y=0.1t$ (${\bm H}\parallel{\bm y}$).
The presence of the remaining ZEP under the Zeeman magnetic field along the edge, which is $H_y$ in this case,
is relevant to the existence of the single Majorana fermion at $k_2>|k_y|>k_1$.
On the other hand, the double TRIMFs at $|k_y|<k_1$ is split into two 
for any direction of Zeeman magnetic field.
Therefore, the height of ZEP $\rho_{s}(0)$ drastically decrease
even if ZEP remains for $\mu_{\rm B}H_y=0.1t$ (${\bm H}\parallel{\bm y}$).

\begin{figure}[htbp]
\begin{center}
\includegraphics[width=7.5cm]{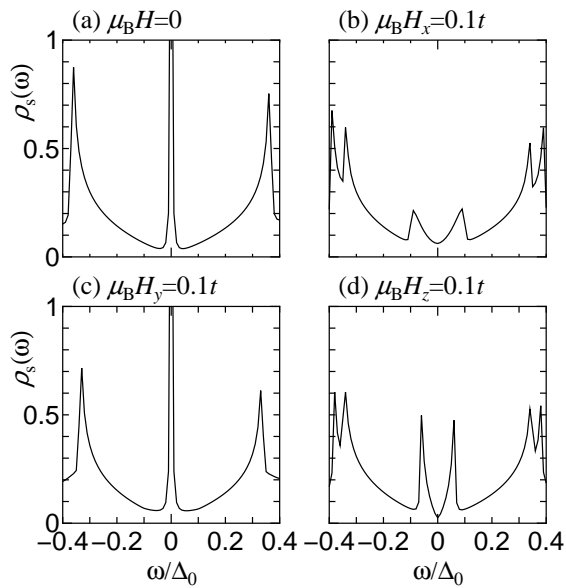}
\caption{  Local density of states at the surface  for $d_{xy}$+$p$-wave 
with $\Delta_{t}=\Delta_{0}$ and 
$\Delta_{s}=0$ for $\lambda=0.5$ and
(a) $\mu_{\rm B}{\bm H}=0$,
(b) $\mu_{\rm B}H_x=0.1t$ (${\bm H}\parallel{\bm x}$),
(c) $\mu_{\rm B}H_y=0.1t$ (${\bm H}\parallel{\bm y}$),
(d) $\mu_{\rm B}H_z=0.1t$ (${\bm H}\parallel{\bm z}$).}\label{fig15}
\end{center}
\end{figure}

Next we consider the triplet dominant $d_{xy}+p$-wave NCS superconductor.
For simplicity, we suppose that only the triplet component $\Delta_{p}$ exists. 
As was shown in Fig.\ref{fig9}(b), in the absence of the Zeeman magnetic
field, a zero energy ABS exists and it can be described
by a single TRIMF.  
The resulting $\rho_{s}(\omega)$ has a sharp ZEP without 
${\bm H}$ [Fig. \ref{fig15}(a)]. 
In a manner similar to the singlet dominant case above, 
the SDOS $\rho_{s}(\omega)$ under the Zeeman magnetic field has a strong
orientational dependence of  ${\bm H}$ as shown in Figs. \ref{fig15}
(b), (c) and (d).  
It is noted that ZEP remains when the applied Zeeman field is along 
$y$-direction. 
This implies that the 
single TRIMF in the $d_{xy}+p$-wave NCS superconductor is robust against 
the Zeeman magnetic field applied in the direction along the edge. 

We would like to emphasis that, as shown from these calculation, the
tunneling spectroscopy with the Zeeman magnetic field is available to
identify the single TRIMF in the $d_{xy}+p$-wave NCS superconductor. 
Simultaneous existence of 
the strong orientational dependence of magnetic field 
and the robust  zero energy peak under a certain direction of the magnetic 
field is a strong evidence to support single Majorana Fermion in this material. 

\section{Summary}
In the present paper, we have 
obtained SDOS, edge modes 
 of NCS superconductors by choosing $s+p$, $d_{x^{2}-y^{2}} + f$, and 
$d_{xy} + p$-wave pair potential based on the lattice model Hamiltonian. 
For $d_{xy} + p$-wave pairing, special ABS appears as a zero energy state 
for $k_{2}>\mid k_{y} \mid>k_{1}$ with wavenumber parallel 
to the interface $k_{y}$, where  $k_{1}$ and $k_{2}$ denote the
magnitude of the Fermi wavenumber in the presence of spin-orbit
coupling. 
The present ABS is a single Majorana edge mode  preserving the time reversal symmetry. 
We have defined a new type of topological invariant number 
and clarify the relevance to the number of 
the time-reversal invariant Majorana edge modes. 
We have found that  
the absolute value of the topological number equals to the number of the 
Majorana edge modes. 
The single Majorana edge mode is generally induced by the 
spin orbit coupling. 
In the presence of the single Majorana edge mode, the SDOS 
has a strong orientational dependence of the 
magnetic field. In a certain direction of the applied magnetic field, 
the single Majorana edge mode is robust and the resulting zero energy peak of 
SDOS remains. 
The present future may serve as a guide to detect Majorana Fermion in 
non-centrosymmetric superconducting superconductors by tunneling spectroscopy.

This work is partly supported by the Sumitomo Foundation (M.S.) and the
Grant-in-Aids for Scientific Research No.\ 22103005 (Y.T. and M.S.), 
No.\ 20654030 (Y.T.) and No.22540383 (M.S.). 


\begin{thebibliography}{99}
\bibitem{dxy}
When the symmetry of the singlet component of pair potential 
is $d_{xy}$-wave ($d_{x^2-y^2}$-wave),
the number of the sign change of the real or imaginary part of 
triplet one on the Fermi surface is two (six).
Thus, we call the mixed pair potential $d_{xy}+p$-wave ($d_{x^{2}-y^{2}} + f$-wave).
\end{thebibliography}
\end{document}